\documentclass[letterpaper,twocolumn,10pt]{article}
\usepackage{usenix2020_SOUPS}
\usepackage{tikz}
\usepackage{amsmath}

\usepackage{filecontents}

\usepackage{cite}
\usepackage{amssymb,amsfonts}
\usepackage{algorithmic}
\usepackage{graphicx}
\usepackage{textcomp}
\usepackage{xcolor}
\def\BibTeX{{\rm B\kern-.05em{\sc i\kern-.025em b}\kern-.08em
		T\kern-.1667em\lower.7ex\hbox{E}\kern-.125emX}}

\usepackage[nolist]{acronym}

\begin{acronym}

    \acro{3DES}{Triple-DES}
    
    \acro{AES}{Advanced Encryption Standard}
    \acro{ALU}{Arithmetic Logic Unit}
    \acro{API}{Application Programming Interface}
    \acro{ARX}{Addition Rotation XOR}
    \acro{ATPG}{Automatic Test Pattern Generation}
    \acro{ASIC}{Application Specific Integrated Circuit}
    \acro{ASIP}{Application Specific Instruction-Set Processor}
    \acro{AS}{Active Serial}
    
    \acro{BDD}{Binary Decision Diagram}
    \acro{BGL}{Boost Graph Library}
    \acro{BNF}{Backus-Naur Form}
    \acro{BRAM}{Block-Ram}
    
    \acro{CBC}{Cipher Block Chaining}
    \acro{CFB}{Cipher Feedback Mode}
    \acro{CFG}{Control Flow Graph}
    \acro{CLB}{Configurable Logic Block}
    \acro{CLI}{Command Line Interface}
    \acro{COFF}{Common Object File Format}
    \acro{CPA}{Correlation Power Analysis}
    \acro{CPU}{Central Processing Unit}
    \acro{CRC}{Cyclic Redundancy Check}
    \acro{CTR}{Counter}
    
    \acro{DC}{Direct Current}
    \acro{DES}{Data Encryption Standard}
    \acro{DFA}{Differential Frequency Analysis}
    \acro{DFT}{Discrete Fourier Transform}
    \acro{DIP}{Distinguishing Input Pattern}
    \acro{DLL}{Dynamic Link Library}
    \acro{DMA}{Direct Memory Access}
    \acro{DNF}{Disjunctive Normal Form}
    \acro{DPA}{Differential Power Analysis}
    \acro{DSO}{Digital Storage Oscilloscope}
    \acro{DSP}{Digital Signal Processing}
    \acro{DUT}{Design Under Test}
    
    \acro{ECB}{Electronic Code Book}
    \acro{ECC}{Elliptic Curve Cryptography}
    \acro{EEPROM}{Electrically Erasable Programmable Read-only Memory}
    \acro{EMA}{Electromagnetic Emanation}
    \acro{EM}{electro-magnetic}
    
    \acro{FFT}{Fast Fourier Transformation}
    \acro{FF}{Flip Flop}
    \acro{FI}{Fault Injection}
    \acro{FIR}{Finite Impulse Response}
    \acro{FPGA}{Field Programmable Gate Array}
    \acro{FSM}{Finite State Machine}
    
    \acro{GUI}{Graphical User Interface}
    
    \acro{HDL}{Hardware Description Language}
    \acro{HD}{Hamming Distance}
    \acro{HF}{High Frequency}
	\acro{HRE}{Hardware Reverse Engineering}
    \acro{HSM}{Hardware Security Module}
    \acro{HW}{Hamming Weight}
    
    \acro{IC}{Integrated Circuit}
    \acro{IO}[I/O]{Input/Output}
    \acro{IOB}{Input Output Block}
    \acro{IOT}[IoT]{Internet of Things}
    \acro{IP}{Intellectual Property}
    \acro{ISA}{Instruction Set Architecture}
    \acro{IV}{Initialization Vector}
    
    \acro{JTAG}{Joint Test Action Group}
    
    \acro{KAT}{Known Answer Test}
    
    \acro{LFSR}{Linear Feedback Shift Register}
    \acro{LSB}{Least Significant Bit}
    \acro{LUT}{Look-up table}
    
    \acro{MAC}{Message Authentication Code}
    \acro{MIPS}{Microprocessor without Interlocked Pipeline Stages}
    \acro{MMIO}{Memory Mapped \acl{IO}}
    \acro{MSB}{Most Significant Bit}
    
    \acro{NASA}{National Aeronautics and Space Administration}
    \acro{NSA}{National Security Agency}
    \acro{NVM}{Non-Volatile Memory}
    
    \acro{OFB}{Output Feedback Mode}
    \acro{OISC}{One Instruction Set Computer}
    \acro{ORAM}{Oblivious Random Access Memory}
    \acro{OS}{Operating System}
    
    \acro{PAR}{Place-and-Route}
    \acro{PCB}{Printed Circuit Board}
    \acro{PC}{Personal Computer}
    \acro{PS}{Processing Speed}
	\acro{PR}{Perceptual Reasoning}
    \acro{PUF}{Physically Unclonable Function}
    
    \acro{RISC}{Reduced Instruction Set Computer}
    \acro{RNG}{Random Number Generator}
    \acro{ROBDD}{reduced ordered binary decision diagram}
    \acro{ROM}{Read-Only Memory}
    \acro{ROP}{Return-oriented Programming}
    \acro{RTL}{Register Transfer Level}
    
    \acro{SAT}{Boolean satisfiability}
    \acro{SEMOBS}{Self-Modifying Bitstreams}
    \acro{SCA}{Side-Channel Analysis}
    \acro{SHA}{Secure Hash Algorithm}
    \acro{SNR}{Signal-to-Noise Ratio}
    \acro{SPA}{Simple Power Analysis}
    \acro{SPI}{Serial Peripheral Interface Bus}
    \acro{SRAM}{Static Random Access Memory}
	
	\acro{VLSI}{Very-Large-Scale Integration}
	
	\acro{WAIS-IV}{Wechsler Adult Intelligence Scale}
	\acro{WM}{Working Memory}
    
    \acro{TRNG}{True Random Number Generator}
    
    \acro{UART}{Universal Asynchronous Receiver Transmitter}
    \acro{UHF}{Ultra-High Frequency}
    
    \acro{vFPGA}{virtual FPGA}
	\acro{VC}{Verbal Comprehension}
    
    \acro{WISC}{Writeable Instruction Set Computer}
    
    \acro{XDL}{Xilinx Description Language}
    \acro{XTS}{XEX-based Tweaked-codebook with ciphertext Stealing}
    
    \end{acronym}
    
\usepackage{csquotes}
\usepackage{xspace}
\usepackage{multirow}
\usepackage{booktabs}
\usepackage{tabularx}
\usepackage{enumerate}
\usepackage{paralist}

\usepackage[normalem]{ulem}
\microtypecontext{spacing=nonfrench}

\usepackage{titlesec}
\titlespacing*{\section} {0pt}{2.5ex plus 1ex minus .2ex}{1.5ex plus .2ex}
\titlespacing*{\subsection} {0pt}{2.25ex plus 1ex minus .2ex}{1ex plus .2ex}
\titlespacing*{\subsubsection} {0pt}{2.25ex plus 1ex minus .2ex}{0.5ex plus .2ex}

\usepackage{float}

\usepackage{pgfplots}
\pgfplotsset{compat=1.9}
\usetikzlibrary{pgfplots.groupplots}

\pgfplotsset{every axis/.append style={
		label style={font=\footnotesize},
		tick label style={font=\footnotesize}  
}}
\usetikzlibrary{arrows.meta}
\tikzset{%
	>={Latex[width=2mm,length=2mm]},
	base/.style = {rectangle, rounded corners, draw=black,
		minimum width=4cm, minimum height=0.75cm},
	process/.style = {base, minimum width=2.5cm, fill=white!15},
}

\definecolor{teal}{HTML}{008081}
\definecolor{turkishblue}{HTML}{4F97A3}
\definecolor{pigeon}{HTML}{7285A5}
\definecolor{powder}{HTML}{B0DFE5}
\definecolor{maya}{HTML}{73C2FB}
\definecolor{cornflower}{HTML}{6593F5}
\definecolor{azure}{HTML}{0080FF}
\definecolor{tiffany}{HTML}{81D3D0}
\definecolor{turquoise}{HTML}{3FE0D0}
\definecolor{olympique}{HTML}{008ECC}

\definecolor{c1}{HTML}{49868C}
\definecolor{c2}{HTML}{A0D9D9}
\definecolor{c3}{HTML}{D9B166}

\definecolor{c4}{HTML}{1AFF1A}
\definecolor{c5}{HTML}{1A1AFF}

\makeatletter
\def\endthebibliography{%
	\def\@noitemerr{\@latex@warning{Empty `thebibliography' environment}}%
	\endlist
}
\newcommand*\bigcdot{\mathpalette\bigcdot@{.8}}
\newcommand*\bigcdot@[2]{\mathbin{\vcenter{\hbox{\scalebox{#2}{$\m@th#1\bullet$}}}}}
\makeatother

\newcommand{\HAL}{\textnormal{HAL}\xspace}

\makeatletter
\newcommand*{\radiobutton}{%
	\@ifstar{\@radiobutton0}{\@radiobutton1}%
}
\newcommand*{\@radiobutton}[1]{%
	\begin{tikzpicture}
	\pgfmathsetlengthmacro\radius{height("X")/2}
	\draw[radius=\radius] circle;
	\ifcase#1 \fill[radius=.6*\radius] circle;\fi
	\end{tikzpicture}%
}
\makeatother

\begin{document}

\date{}

\title{\Large \bf An Exploratory Study of Hardware Reverse Engineering\\Technical and Cognitive Processes}

\def\plainauthor{Steffen Becker, Carina Wiesen, Nils Albartus, Nikol Rummel, Christof Paar}

\author{\rm{Steffen~Becker$^*$$^\dagger$$^\ddagger$, Carina~Wiesen$^*$$^\dagger$$^\ddagger$, Nils~Albartus$^\dagger$$^\ddagger$, Nikol~Rummel$^\dagger$, and~Christof~Paar$^\ddagger$}\\
	\normalsize{$^\dagger$\textit{Ruhr University Bochum}; $^\ddagger$\textit{Max Planck Institute for Cybersecurity and Privacy}}\\
	\normalsize{\{steffen.becker, carina.wiesen, nils.albartus, nikol.rummel\}@rub.de}; christof.paar@csp.mpg.de\\
	\footnotesize{$^*$Both authors contributed equally to this work.}
	}

\maketitle
\thecopyright

\begin{abstract}
Understanding the internals of \acp{IC}, referred to as \ac{HRE}, is of interest to both legitimate and malicious parties.
\ac{HRE} is a complex process in which semi-automated steps are interwoven with human sense-making processes.
Currently, little is known about the technical and cognitive processes which determine the success of \ac{HRE}.
	
This paper performs an initial investigation on how reverse engineers solve problems, how manual and automated analysis methods interact, and which cognitive factors play a role.
We present the results of an exploratory behavioral study with eight participants that was conducted after they had completed a 14-week training.
We explored the validity of our findings by comparing them with the behavior (strategies applied  and solution time) of an \ac{HRE} expert. 	
The participants were observed while solving a realistic \ac{HRE} task.
We tested cognitive abilities of our participants and collected large sets of behavioral data from log files.
By comparing the least and most efficient reverse engineers, we were able to observe successful strategies.
Moreover, our analyses suggest a phase model for reverse engineering, consisting of three phases.
Our descriptive results further indicate that the cognitive factor \ac{WM} might play a role in efficiently solving \ac{HRE} problems.
Our exploratory study builds the foundation for future research in this topic and outlines ideas for designing cognitively difficult countermeasures (\enquote{cognitive obfuscation}) against \ac{HRE}. 
	
\end{abstract}
\section{Introduction}
By definition, every computing system is based on hardware components, in particular on \acfp{IC}. Their internals are typically completely opaque to the user and largely even to the developers of the system. Understanding the internals of (digital) hardware components, which requires \acf{HRE}, is of interest for both malicious and legitimate reasons \cite{JETC:2016:Quadir}. For instance, the sensitive topic of low-level backdoors (i.e., hardware Trojans), which underlies the current discussion about foreign-built communication and computer equipment \cite{robertson_big_2018, satariano_huawei_2019},  requires \ac{HRE} for detection of such manipulations. On the other hand, adversaries might also need to reverse engineer the hardware they plan to subvert. \ac{HRE} is also widely-used in practice for detection of \ac{IP} infringements \cite{ivsw:2017:fyrbiak}. On the adversary side, a malicious party needs to reverse areas of interest or even entire \acp{IC}. Moreover, there is a host of Trojan detection techniques \cite{dtc:2010:tehranipoor} that require a flawless  model of the target \ac{IC}.

Despite its relevance, \ac{HRE} is relatively poorly understood compared to many other areas of both hardware design and computer security \cite{ivsw:2017:fyrbiak}. 
We argue that it is desirable to obtain a better understanding of \ac{HRE}. 
First, it will aid with assessing the threat posed by adversaries performing \ac{HRE}. This is particularly prudent because there is undoubtedly expertise about \ac{HRE} within government agencies with malicious intent \cite{macaskill_nsa_2013}. Second, \ac{HRE} performed for defensive purposes such as \ac{IC} verification will benefit from having a better understanding of the involved time and costs. Third, having a better grasp on \ac{HRE} will allow us to derive sound obfuscation techniques that exploit cognitive limitations of humans.

\ac{HRE} is a multilayered process, where high-level information is extracted from a low-level circuit, consisting of two major stages \cite{azriel_sok:_2019}: In the first stage, a gate-level netlist is obtained from an \ac{IC} either directly from the device or possibly through (online) interception of design information. A gate-level netlist is a logical circuit description and is usually composed of Boolean gates and their respective interconnections.
The technical steps required for netlist extraction, including decapsulation and imaging, are relatively well-covered in the literature \cite{ches:2009:torrance, JETC:2016:Quadir}. The objective of the second stage is to \textit{make sense} of the recovered netlist, that is, to \textit{understand} the netlist \cite{azriel_sok:_2019}. This second stage has only been rudimentarily addressed in the literature and is the topic of our investigation.

In this contribution, we present an exploratory study with the goal of obtaining initial insights into the complex processes of the sense-making part of \ac{HRE}. This undertaking is particularly challenging as it depends on non-trivial technical steps as well as on cognitive factors of the human reverse engineer. To the best of our knowledge, it is the first time such a study based on observing a group of reverse engineers has been conducted. 


\subsubsection*{Overview on our Exploratory Study}
In order to gain such insights, we conducted an empirical study that explores both the technical and the underlying cognitive processes of hardware reverse engineers. In an ideal scenario, we would examine expert reverse engineers working on real-world tasks with tools they are familiar with. However, we face the methodological problem that those experts are few to begin with. They are primarily active in government agencies and a few highly specialized companies, and are generally unavailable to the scientific community. We approach this problem by observing students enrolled in a 5-years BSc-MSc program in cyber security while solving a realistic \ac{HRE} task, which was developed in collaboration with \ac{HRE} experts. Prior to the study, the students had been exposed to an extensive 14-week \ac{HRE} training. For the study, we selected eight students based on their performance during the \ac{HRE} training.

Despite the difficulty of engaging \ac{HRE} experts as study participants, we were able to recruit one expert via the professional network of one of the authors.
This expert solved the same task as our participants.
The collected data served as a sanity check for our student sample with respect to solution time, as well as phases of and applied strategies during \ac{HRE}.

In our exploratory study, we collected detailed behavioral data during an \ac{HRE}-based attack and measured cognitive factors of eight participants. The \ac{HRE} task is based on a realistic setting, where the analyst has to circumvent an \ac{IP} protection mechanism in an unknown circuit. Analysis of the data revealed initial insights into problem solving strategies and relevant cognitive factors in \ac{HRE}. We were also able to derive first hypotheses for a novel class of obfuscation measures that take the boundaries of cognitive abilities into account.
Although our work was faced with methodological challenges that we discuss in the limitations section, we render the following main contributions advancing the current scientific knowledge of technical and cognitive processes in \ac{HRE}.
\begin{compactenum}[1)]
	\item We propose and examine an \ac{HRE} phase model based on an exploratory behavioral study of human reverse engineers, who solved a realistic \ac{HRE} task.
	\item Based on behavioral analyses we explore more and less efficient \ac{HRE} strategies of reverse engineers.
	\item We explore the role of different cognitive factors in efficiently solving \ac{HRE} problems.
	\item Based on our findings, we derive hypotheses for a novel class of \ac{HRE} countermeasures called cognitive obfuscation and outline future research directions. 
\end{compactenum}

\section{Background}
In this section, we present the relevant technical background of \ac{HRE} and propose a three-phase model encompassing human processes during \ac{HRE}. Furthermore, we introduce related work on cognitive processes in reverse engineering. Against this background, we identify the research gap and derive research questions we seek to answer in this work.

\subsection{Hardware Reverse Engineering} \label{ssec:background_HRE}
Reverse engineering is the process of extracting knowledge or design information from anything man-made in order to comprehend its inner structure \cite{smc:1985:rekoff}. As mentioned above, in the case of \ac{HRE}, there are two distinct stages \cite{azriel_sok:_2019}. In the first stage, a gate-level netlist is obtained directly from an \ac{IC} or a \ac{FPGA} or through (online) interception of design information.
Although netlist extraction requires sophisticated technical methods, research has shown that netlists can be extracted reliably by trained specialists from both, \acp{IC} and \acp{FPGA} \cite{ches:2009:torrance, mm:2013:ding, ccs:2011:moradi}.

In the second, sense-making stage of \ac{HRE}, the netlist is transformed into higher levels of abstraction that enable a detailed analysis. This often involves module recognition, identification of blocks of interest, and detailed understanding of Boolean sub-circuits \cite{tect:2013:subramanyan, ches:fsm:2018, azriel_sok:_2019}. 
The analysis typically serves a specific objective, for example, finding and understanding \ac{IP} blocks or extraction of cryptographic keys \cite{ivsw:2017:wallat}.
Due to the nature of this stage --- which requires human ingenuity, sense-making, and in many cases customized solutions --- fully automated tools do not exist\cite{ivsw:2017:fyrbiak}. Instead, the analyst typically employs \ac{HRE} tools which enable interaction with the target netlist. Tools may provide semi-automated support for the human analyst, for example, for running specific algorithms on the netlist \cite{chisholm1999understanding}, as well as, features for manual analysis of netlist components\cite{wallat_highway_2019}.


Even with tool support, the cognitive processes and human problem-solving strategies are crucial for \ac{HRE}, yet remain poorly understood. Against this background, it is hardly surprising that hardware obfuscation, which is a widely used countermeasure against \ac{HRE}, is largely based on ad-hoc methods \cite{wiesen2019towards}. We argue that a comprehension of human processes in \ac{HRE} will open a pathway for the development of novel obfuscation techniques.
This paper will conclude with first guidelines for how such cognitive obfuscation measures might look like. 
Reverse engineering of large and complex netlists is commonly driven by an objective more narrow than full understanding of the entire netlist.
We model a situation under which \ac{HRE} is typically performed in practice through the following conditions:
\begin{compactitem}
	\item An (error-free) netlist of the entire design or the area of interest is at hand.
	\item An \ac{HRE} tool is available that allows interaction with the target netlist.
	\item There is a clear objective, for example, removal of an \ac{IP} protection mechanism.
	\item (First) hypotheses related to the objective exist, for example, which \ac{IP} protection mechanism is implemented.
\end{compactitem}
If these preconditions are met, a human analyst attempts to understand the target netlist through two principal means: 
\begin{compactitem}
	\item \textbf{Manual analyses}: Detailed manual inspection of netlist components and explorative navigation through textual and graphical representations of the netlist.
	\item \textbf{Semi-automated analyses}: Customized scripts and programs that facilitate structural and functional analyses of the  netlist.
\end{compactitem}
In Section~\ref{sec_Results}, we will explore which role these two mechanisms play during \ac{HRE}.

\subsection{Phase Model for Gate-level Netlist Reverse Engineering}
\label{phasemod}
Due to the sheer complexity of the sense-making stage, it is plausible that human strategies during this process can be divided into sub phases of human sense-making.
Thus, we propose a three-phase model derived from several \ac{HRE}-based attacks from literature \cite{ches:fsm:2018, azriel_sok:_2019}, and two technical \ac{HRE} workflow descriptions \cite{chisholm1999understanding, tect:2013:subramanyan}. Even though not explicitly formulated in a model before, the three phases are an implicit hypothesis about the inner workings of \ac{HRE}.
In a very recent work, Votipka et al. introduce a similar model of human processes during software reverse engineering based on expert interviews \cite{votipkaobservational}.
In the following passages, we briefly characterize each phase under the assumption that the preconditions described above are fulfilled.
Also, since netlist reverse engineering is a continuous process, the consecutive phases can blend into each other rather than being strictly disjoint.

\subsubsection*{Phase 1: Candidate Identification}
The goal of Phase~1 is to identify single candidates and subcircuits which are potentially relevant in the context of the reverse engineering process. Therefore, the analyst starts exploration via structural analyses of the netlist topology with the goal to identify blocks of interests. The result are sub-circuit candidates, which are further inspected in Phase~2.
Suited methods for this phase are semi-automated structural analyses (e.g., graph clustering algorithms or sub-circuit matching) as well as custom-tailored structural analyses incorporating hypotheses about searched structures. Additionally, manual netlist exploration can provide a starting point for the reverse engineer, for example, by inspecting global in- and outputs.

\subsubsection*{Phase 2: Candidate Verification}
The goal of Phase~2 is the verification of extracted candidates from Phase~1 in order to narrow them down and select target components for Phase~3. If no target components are remaining, new candidates have to be identified in Phase~1 iteratively by refining the methods used.
Phase~2 incorporates static analyses methods such as Boolean functionality analysis or sub-circuit matching. Manual inspection of candidates can support the reverse engineer by testing the existing hypotheses before solving the problem algorithmically.

\subsubsection*{Phase 3: Realization}
While Phases~1 and~2 can be generalized for most \ac{HRE} tasks, the goals and procedures of Phase~3 differ significantly depending on the objective. 
Methods employed include sub-circuit interpretation and annotation in order to obtain a more abstract netlist model; netlist simulation to analyze sequential behavior; or preparation of malicious netlist manipulation, for example, add, remove, or change functionality of netlist components. In many cases, Phase~3 incorporates and combines several of the aforementioned methods.

\subsection{Cognitive Processes in Reverse Engineering}
As outlined above, \ac{HRE} always involves sense-making processes. Thus, the success of \ac{HRE} heavily depends on skills, knowledge, and expertise of the performing reverse engineer. Surprisingly, underlying cognitive processes in \ac{HRE} are understudied and remain poorly understood \cite{ivsw:2017:fyrbiak}. Despite this observation, prior research on \ac{HRE} almost solely focuses on technical factors. Nonetheless, one prior work explored cognitive processes by defining reverse engineering of Boolean systems as a specific type of human problem solving \cite{lee2013theory}. In general, a problem exists when a person lacks in knowledge which enables the problem solver to achieve a desired goal \cite{fischer2011process}. Problem solving is defined as a sequence of cognitive operations (e.g., problem solving strategies) in order to solve a task for which the individual does not possess a suitable routine operation \cite{newell1972human}. The success of problem solving is influenced by several factors, for example, prior domain specific knowledge \cite{beckmann2014benefit}, or cognitive abilities (e.g., intelligence and sub factors like \acl{WM}) \cite{hambrick2003role}. Baddeley demonstrated that brain systems like the \acl{WM} are essential in solving complex cognitive tasks like problem solving \cite{baddeley1992working}. Besides cognitive abilities, the level of expertise \cite{chase1973perception} is another important factor in determining problem solving performances. Larkin et al. showed that experts were quicker in solving physics problems than novices \cite{larkin1980expert}.

In 2013, Lee and Johnson-Laird analyzed problem solving behavior in reverse engineering of Boolean systems by conducting five experiments in a laboratory setting \cite{lee2013theory}. The tasks students were asked to solve merely involved drawing Boolean circuits controlling an electric light. Consequently, the ecological and external validity of these experiments appears low in the context of \ac{HRE}.
Thus, it remains unclear to what extent the results of Lee and Johnson-Laird can be generalized to reverse engineering of entire \acp{IC}, which commonly consist of hundred of thousands or millions of logic components. Nevertheless, we transfer human problem solving processes to the cognitive processes in \ac{HRE} by observing problem solving strategies of more and less successful reverse engineers and by measuring cognitive factors which might play a role in \ac{HRE} problem solving. 

\subsection{Research Gap and Research Questions}
\label{subsec:rq}
In this work, we aim to close the existing research gap by providing first insights into the technical and cognitive processes during a realistic \ac{HRE} task. We qualitatively examine the occurrence of the 3-phase model for netlist reverse engineering in the participants' behavior, and investigate \ac{HRE} problem solving strategies and their influencing cognitive factors on the basis of behavioral and cognitive data. This allows us to derive hypotheses for cognitive obfuscation techniques, i.e., novel countermeasures impeding \ac{HRE}. The paper at hand attempts to answer the following research questions:\vspace{.1cm}\\
\textbf{RQ1.} Can the phases of human sense-making be detected during \ac{HRE} processes? If so, which are the crucial phases?\vspace{.1cm}\\ 
\textbf{RQ2.} Which strategies distinguish more and less efficient reverse engineers?\vspace{.1cm}\\ 
\textbf{RQ3.} Which cognitive prerequisites play a role for the success of \ac{HRE}?\vspace{.1cm}\\
\textbf{RQ4.} How can those insights be used to derive hypotheses for cognitive obfuscation?\vspace{.1cm}\\ 
Our exploratory study opens many venues for further in-depth research on the challenging interdisciplinary problem of understanding \ac{HRE}.
\section{Methodology}
In the following section, we first describe the research environment enabling our study.
Second, we outline the details of our user study, including our participants and all study related measures and processes.
Last, we explain the behavioral analyses providing the underlying data for our exploration of human factors in \ac{HRE}.

\subsection{Research Environment}
This section outlines the research environment consisting of an extensive \ac{HRE} training, the \ac{HRE} tool \HAL, and a practical \ac{HRE} task under study.
\subsubsection{\ac{HRE} Training}
\label{subsub:course}
The contents of the 14-week \ac{HRE} training were developed based on the comprehensive body of technical research and industry practices and on educational guidelines facilitating learning \ac{HRE} with input from experts from academia and practice \cite{TALE:2018:Wiesen}.
Subsequently, it was shown that the training successfully promotes \ac{HRE} skill acquisition \cite{wiesenpromoting}. 
During the first six weeks, the instructors conveyed necessary theoretical backgrounds, before students solved four training tasks with the \ac{HRE} tool \HAL in the 8-week practical part.
The training was followed by a two-week study, where participants solved a realistic \ac{HRE} task.
Crucially, neither the proposed \ac{HRE} phase model, nor the concrete scenario of, or any solution strategies for the Study Task were part of the training.


\subsubsection{\ac{HRE} Tool HAL}
Given today's integration density, which commonly results in very complex netlists, it is virtually impossible to reverse even moderately sized \ac{IC} or \ac{FPGA} designs without tool support. Therefore, it is safe to assume that professional reverse engineers, for example, in government agencies and specialized companies, have access to such (internally developed) tools, cf. \cite{macaskill_nsa_2013, thomas_texplained_nodate}. Recently, the open-source framework \HAL \cite{wallat_highway_2019, tdsc:2018:fyrbiak} has become available on GitHub, which is the first tool specifically designed to facilitate \ac{HRE}. By using \HAL in our study, we create an environment that is, thus, similar to one encountered in real-world \ac{HRE} situations.

\HAL operates on gate-level netlists. It has a modular and extendable design and supports both, static analysis and cycle-accurate simulation of netlists. There are several references describing semi-automated reverse engineering and manipulation tasks using \HAL \cite{ivsw:2017:wallat, tdsc:2018:fyrbiak, ches:fsm:2018}. 
Crucially for our study, \HAL can capture all user interactions and therefore enables the investigation of the many sub-steps that take place during \ac{HRE}, including the study of human factors. Below is a description of \HAL features that were particularly  useful for our  study:
\begin{compactitem}
	\item \HAL offers an \textit{interactive GUI} allowing detailed manual inspection and exploration of netlist components as well as module grouping functionalities.
	\item \HAL natively implements a \textit{Python interface} enabling script-based interactions with the netlist.
	\item \HAL comes with a variety of \textit{accompanying materials}, most importantly a detailed documentation of \HAL-specific Python commands and a coding guide providing many examples for common use cases.
\end{compactitem}

\subsubsection{Tasks and Materials}
\label{ssec:tasks}
In the practical part of the \ac{HRE} training and in the study, participants worked on five reverse engineering tasks based on flat netlists synthesized for \acp{FPGA}. Those netlists do not contain any high-level information about the design such as component names, module boundaries or hierarchy elements. Netlist components are composed of \acp{LUT} with up to six inputs, which realize the circuit's Boolean functions, multiplexers, \acp{FF}, and their respective logic interconnections.

All five tasks are based on ideas drawn from recent literature on \ac{HRE} attacks and countermeasures \cite{azriel_sok:_2019,ches:fsm:2018,tcad:2009:chakraborty}. Thus, they represent a number of different \ac{HRE} settings ranging from sub-circuit detection over obfuscation circumvention up to malicious manipulation of cryptographic cores. The tasks are designed with increasing difficulty, that is, the problem itself increased in complexity, the level of guidance decreased, and the size and complexity of the target netlist increased, cf. Table~\ref{tab1:tasks}.
The level of guidance includes task-related support by instructors, which participants received only during the training tasks, as well as the amounts of accompanying materials, e.g., (excerpts of) scientific papers, or relevant examples from the coding guide for the task.
\begin{table}[htbp]
	\small
	\begin{center}
		\caption{Difficulty, level of guidance, and netlist complexity for the \ac{HRE} tasks on a scale from \textbf{+} (low) to \textbf{+++} (high).}
		\label{tab1:tasks}
		\begin{tabular}{rccc}
			\toprule
			                & \textbf{\iffalse Task\fi Difficulty} & \textbf{\iffalse Level of \fi Guidance} &       \textbf{\iffalse Netlist\fi Complexity}        \\ \cmidrule(lr){1-4}
			\textbf{Training Task 1} &        \textbf{+}        &        \textbf{+++}        &  \textbf{+} \\ \cmidrule(lr){2-4}
			\textbf{Training Task 2} &       \textbf{++}        &        \textbf{+++}        &  \textbf{+} \\ \cmidrule(lr){2-4}
			\textbf{Training Task 3} &       \textbf{++}        &        \textbf{++}         &  \textbf{+} \\ \cmidrule(lr){2-4}
			\textbf{Training Task 4} &       \textbf{+++}       &        \textbf{++}         & \textbf{++} \\ \cmidrule(lr){1-4}
			\textbf{Study Task} &       \textbf{+++}       &         \textbf{+}         & \textbf{++} \\
			\bottomrule
		\end{tabular}
	\end{center}
\end{table}

In the following, the last and most difficult task, which provides the data for our study, is introduced.
\vspace{.2cm}\\
\textbf{Study~Task: Breaking Watermarkings.}\label{task5} In this task, the \ac{IP} protection mechanism---a so-called watermarking scheme---proposed by Schmid et al. \cite{fpt:2008:schmid} is embedded in a hardware design, enabling the detection of \ac{IP} theft.
The analyst has the objective to clone the underlying netlist without copying its watermark, and therefore thwarting detection of \ac{IP} theft in the cloned circuit.
While materials such as the original paper on the functionality of the watermarking scheme were provided, the analysts had to develop the strategies for detection and removal of the watermarking themselves.

To enable analysis of \ac{HRE} strategies employed by the participants while solving the Study Task, netlist components are pre-annotated in relevant and irrelevant ones.
Out of 4653 total netlist components, 54 were marked as relevant since they are associated with the watermark. 50 of those components are \textit{candidates}, thereof 30 actually implement the watermark and are therefore \textit{targets}. The remaining 4 components are important anchor points for the watermark detection. Naturally, the annotation was invisible to the participants. All netlist components have a unique identifier in \HAL, which allows their tracing during the behavioral analyses described in Section~\ref{subsubsec:behavanalysis}. 


\subsection{User Study and Variables}
This section describes our user study in detail. This includes the specification of our study participants and the expert, ethical considerations concerning this research, as well as an outline of the study procedures and variables.

\subsubsection{Study Participants}
We conducted our quasi-experimental study with a within-subject design, where every participant had to solve the Study Task using \HAL. Since experts were generally unavailable for this research, we approximated experienced reverse engineers by recruiting senior-level and MSc-level students with a (target) degree in cyber security. The recruited participants acquired relevant \ac{HRE} knowledge and skills during the extensive training phase and were selected based on their performance in the four training tasks described in Section~\ref{subsub:course}. Furthermore, we approximated the situation of experienced reverse engineers by providing a coding guide containing relevant code snippets, instructions, and programming methods which were used throughout the study and the training tasks. 

Overall, 22 participants volunteered to participate in the study.
All participants were enrolled in either the last year of a three-year bachelor's or in a master's program in cyber security. 
Five of the original 22 participants decided to drop out.
Furthermore, three participants were excluded due to incomplete datasets.
Out of the remaining 14, eight participants (mean age: $24$ years; $SD = 4$ years; one graduate student) were selected based on their performance in the training tasks, i.e., the average solution probability of all training tasks was above $95\%$ per selected participant.
Solution probabilities were evaluated on a scale from $0\%$ to $100\%$ by three teaching assistants based on a detailed gradebook with sample solutions.

\subsubsection{Sanity Check with Expert}
Despite the difficulty of recruiting \ac{HRE} experts as study participants, we were able to recruit one expert through the professional network of one of the authors.
The expert also performed the Study Task. 
There was a two-fold objective to engaging the expert. 
First, the comparison of the reversing behavior between the expert and the study participants allowed us to explore if the students' approximated level as experienced reverse engineers was an adequate assumption. Second, we could evaluate the difficulty of the Study Task.
For assessing the status as \ac{HRE} expert, we followed the criteria of Votipka et al. \cite{votipkaobservational}:  Our expert had $5$ years of experience and a self-assessed skill-level of  $4$ on a 5-point Likert-Scale (with $1$ being a beginner and $5$ being an expert).
The expert gave written informed consent on using the collected socio-demographic and behavioral data in the context of this research project.
Cognitive abilities were not tested for the expert to protect the person's privacy in case of a potential de-anonymization.

The expert received the underlying netlist of the Study Task and corresponding materials (virtual machine running \HAL, coding guide, documentation of Python commands, and paper on the implemented watermarking) and was asked to remove the \ac{IP} protection mechanism from the circuit. Moreover, the expert was already experienced in working with \HAL.
Subsequently, the behavioral data collected while the expert was working on the Study Task was analyzed in order to compare the observations made for our student sample with respect to solution time, as well as phases of and applied strategies during \ac{HRE} with the expert's problem solving behavior.

\subsubsection{Control Variables}
\label{subsubsec:control}
We asked participants to provide information about their socio-demographics (age, major, target degree, number of semesters enrolled) in a self-developed questionnaire (cf. Appendix~\ref{appendix:questionnaires}). 

\subsubsection{Cognitive Abilities}
\label{subsubsec:cognitive}
Based on the definition of reverse engineering as a specific type of problem solving \cite{lee2013theory}, we transferred the concept of problem solving research into the domain of \ac{HRE}. Therefore, as problem solving performances can depend on cognitive abilities \cite{hambrick2003role}, we measured sub factors of general intelligence and their correlations to \ac{HRE} problem solving performances. As a measure for problem solving performance, we correlate the variable time on task, which is a traditional measure in cognitive psychology, with levels of cognitive abilities. In order to assess the participants’ levels of cognitive abilities, sub tests of a valid test instrument, the \ac{WAIS-IV} \cite{wechsler2008wechsler}, were used. We integrated the following three sub scores in our study: \ac{PR}, \acf{WM}, and \ac{PS}. The fourth test of the \ac{WAIS-IV} on \ac{VC}, which measures verbal reasoning and verbal expression, was not included in the study since participants had different native languages.
\ac{PR} measures the ability to accurately interpret and work with visual information. The sub score \ac{WM} assessed the ability to store information and to perform mental operations on that stored information. The third score \ac{PS} reflected the ability of processing visual information quickly and efficiently.

\subsubsection{Ethical Considerations}
Our institute does not have an ethics board or IRB, but the study protocols were reviewed and approved by the universities' data protection officer.
Before entering the \ac{HRE} training, all 22 participants gave written informed consent. They received monetary compensation well above minimum wage levels for time spent on materials related to our study.
We informed participants that they can withdraw from our study at any time and that all partial data will not be analyzed or stored.
Privacy was ensured by randomly assigning pseudonyms to the participants, which were used instead of their actual names throughout all materials related to our study.

\subsubsection{Study Procedures}
The \ac{WAIS-IV} was conducted in a 60 to 90 minutes face-to-face session with each participant prior to the study.
Before starting the Study Task, participants were asked to answer the questionnaire on socio-demographics via the online survey provider Soscisurvey. 
After finishing the Study Task, participants uploaded log files recorded while solving the Study Task onto a server located at the university.

\subsection{Behavioral Analyses}
\label{subsubsec:behavanalysis}

In order to explore human processes in \ac{HRE}, we analyzed log files automatically generated by \HAL.
We collected one log file --- containing several thousands up to tens of thousands of log entries --- for each participant solving the Study Task.
Every log entry consists of a timestamp and one of the following events in \HAL: (i) content and terminal output of the executed Python script; (ii) manual selection of netlist component (unique identifier) via GUI; or (iii) additional system-level entries such as indications for user (in)activity. 
The following parameters were extracted from the collected logfiles and serve as the basis for the behavioral analyses.\vspace{.1cm}\\
\textbf{Total Solution Time.} Since participants solved the task over a period of two weeks, relative solution time was calculated, where loading the inspected netlist for the first time in \HAL marks the starting time.
All periods lasting more than 10 minutes without log entries were manually reviewed. Based on observed wall clock time and script changes during those periods, a threshold of 60 minutes was identified for periods of inactivity, which were subsequently excluded from time on task (cf. Table~\ref{tab:inacticity} in Appendix~\ref{appendix:results}).\vspace{.1cm}\\
\textbf{Executed Scripts.}
Executed scripts were saved as standalone versions together with their associated execution time and accessed netlist components.
Each script execution represents an intermediate state of an iteratively developed solution script.\vspace{.1cm}\\
\textbf{Manual Component Selections.}
Manual selections of netlist components together with the associated time were extracted.\vspace{.1cm}\\
\textbf{Time per Phase.}
The phases of human sense-making as described in Section~\ref{phasemod} were detected by checking the following conditions for the completion of each phase:\\
\textit{Phase~1} is completed when all 50 candidates are identified in the netlist. Thus, they have to be accessed by a script.\\
\textit{ Phase~2} is finalized when all 30 targets which actually implement a watermark are identified and the watermarking signatures are read out. Therefore, data from those components has to be extracted and processed via script.\\
\textit{Phase~3} is finished after removing all watermark signatures from the circuit. This implies that all 30 targets have to be manipulated. Consequently, the script solving Phase~3 has to conduct a manipulation effectively removing the watermarking on those targets.

Fulfillment of those conditions was checked through a combination of automated script analysis, for example, when and how relevant components are accessed, as well as further manual reviews of scripts. Manual reviews were conducted collaboratively by two researchers familiar with the task under study.\vspace{.1cm}\\
\textbf{Progress Metric.}
As described in Section~\ref{ssec:background_HRE}, reverse engineers use the two principal actions of semi-automated scripts and manual analysis.
We assessed progress during the Study Task with respect to both actions.
\vspace{.1cm}\\\textit{Script Progress Score.} Regarding semi-automated reverse engineering, we employed automated script analysis, e.g, the observation of persistent lines of code with respect to the final solution script, and further manual reviews of single script iterations.
For each executed script, a progress score between $0$ and $3$ was assigned according to the rules below:
\begin{compactitem}
	\item[0:] no progress made: no or only obsolete code added,
	\item[1:] few relevant line(s) of code implemented; small subproblem solved,
	\item[2:] relevant block of code or single significant line of code implemented; important subproblem solved,
	\item[3:] significant idea or significant block of code implemented.
\end{compactitem}
\vspace{.1cm}\textit{Manual Progress Score.} With respect to manual analysis, groups of manual interactions were assessed on the same scale by first pooling consecutive component selections and then allocating the following progress scores between~$0$ and~$3$:
\begin{compactitem}
	\item[0:] selection of irrelevant component(s),
	\item[1:] repeated selection of a single relevant component,
	\item[2:] repeated selection of multiple relevant components,
	\item[3:] first selection of relevant component(s).
\end{compactitem}
\vspace{.1cm}\textit{Progress Visualization.} To enable visual representations of progress during \ac{HRE} processes, we allocated weights to the progress scores. Progress Scores~0 and 1 were weighted as is, while Scores 2 and 3, which represent the main progress and occur only sparsely, were allocated a weight of 3 respectively 7. Those weighted scores served as foundation for the graphical progress visualization as follows: For each participant and phase, all assigned weighted scores were summed up. Progress made is represented as the fraction of the weighted progress score divided by the aforementioned sum.
\section{Results}
\label{sec_Results}
In the following section, we briefly illustrate the results of the expert and classify them in relation to the results of our eight participants before we examine the behavioral data and cognitive abilities of the participants with respect to the research questions formulated in Section~\ref{subsec:rq}.
Section~\ref{subsec:empob} provides insights for RQ1 by analyzing occurrence and relevance of phases from the three-phase model (cf. Section~\ref{phasemod}).
With regard to RQ2, the respective strategies of the overall most and least efficient participants are evaluated in Section~\ref{casestud}.
Section~\ref{cognitiveanalysis} presents first indications about cognitive factors and \ac{HRE} regarding RQ3.

\subsection{Sanity Check with \ac{HRE} Expert}
\label{subsec:expertresults}
The total solution time of the recruited expert for correctly solving the Study Task in \HAL was 162~minutes, which is comparable to the fastest participant of our study (169~minutes). The occurrence of all three phases could clearly be detected via analysis of the behavioral data as described in Section~\ref{subsubsec:behavanalysis} (cf. Figure~\ref{fig:solutiontime}). This is notable because the expert has an entirely different training background than the study participants. 
In Phase~1, the expert identified the watermark candidates and saved them in a data structure, before writing a function to successfully extract and functionally verify the watermark signatures, therefore identifying the targets. In the last phase, the expert removed the watermark signatures and generated the watermark-free netlist.

With 164~executed scripts, the expert lies within the upper range of our participants, who executed between 62~and 175~scripts (cf. Appendix~\ref{appendix:results}, Table~\ref{tab:scripts}). Also, the number of 204~manual component selections is comparable to our participants' (cf. Appendix~\ref{appendix:results}, Table~\ref{tab:gui}).

Furthermore, the in-depth analysis of \ac{HRE} strategies revealed that the expert applied strategies similar to those of the fastest participant (e.g., divide-and-conquer), and only rarely ran into periods of stagnation (cf. Figure~\ref{fig:strategy2} in Appendix~\ref{appendix:results}).

In summary, the expert's results and behavior are in line with the study participants, and therefore 
allow the assumption 
of classifying them as experienced reverse engineers. At the same time, the \ac{HRE} expert solved the Study Task in a time comparable to the participants' time on task, which indicates the adequate difficulty of the task.


\subsection{Empirical Observation of Phases (RQ1)}
\label{subsec:empob}
All participants were able to solve the Study Task correctly as indicated by solution probabilities between $97\%$ and $100\%$.
Based on the behavioral analysis described in Section~\ref{subsubsec:behavanalysis}, the occurrence of the three phases proposed in Section~\ref{phasemod} could be observed for every participant. First, participants saved the watermark candidates in (different) data structures in Phase~1, before iterating over those candidates to apply their specific methods of functional verification to identify the targets in Phase~2. Only after successful verification, the participants began to remove the watermarking and to subsequently generate the watermark-free netlist in Phase~3.
In the following passages, we report our results regarding spent time, executed scripts, and manually inspected netlist components per phase.

\subsubsection{Solution Time per Phase}
\begin{figure*}[htpb]
	\begin{tikzpicture}
	\begin{axis}[
	xbar stacked,
	xmin=0, xmax=405,
	width=\linewidth,
	bar width=0.3cm,
	height=7cm,
	ylabel={Participant},
	ytick=data,
	yticklabels from table={figures/graphs/bar_graph_time.csv}{Name},
	xtick= {0 , 60, 120, 180, 240, 300, 360, 420},
	xlabel={Time in Minutes},
	legend entries={Phase 1,Phase 2, Phase 3},
	y tick label style={/pgf/number format/.cd,%
		scaled y ticks = false,
		set thousands separator={},
		fixed
	},]

\addplot+ [c1, fill=c1!100 , point meta={round(((\thisrow{Phase1} / \thisrow{Total})) *100)}, nodes near coords={\textcolor{olympique!10!black}{\small\pgfmathprintnumber\pgfplotspointmeta\%}}] table [y expr=\coordindex, x={Phase1}] {figures/graphs/bar_graph_time.csv};
\addplot+ [c2, fill=c2!100, point meta={round(((\thisrow{Phase2} / \thisrow{Total})) *100)}, nodes near coords={\textcolor{teal!10!black}{\small\pgfmathprintnumber\pgfplotspointmeta\%}}] table [y expr=\coordindex, x={Phase2}] {figures/graphs/bar_graph_time.csv};
\addplot+ [c3, fill=c3!100, point meta={round(((\thisrow{Phase3} / \thisrow{Total})) *100)}, nodes near coords={\textcolor{turkishblue!10!black}{\small\pgfmathprintnumber\pgfplotspointmeta\%}}] table [y expr=\coordindex, x={Phase3}] {figures/graphs/bar_graph_time.csv};

\end{axis}
\draw[dashed] (0,4.7) -- (14.13,4.7);
\draw[dashed] (15.9,4.7) -- (16.2,4.7);
\end{tikzpicture} 
\caption{Absolute and relative solution times per participant and phase. For comparison, solution times of the expert are represented above the dotted line.}
\label{fig:solutiontime}
\end{figure*}
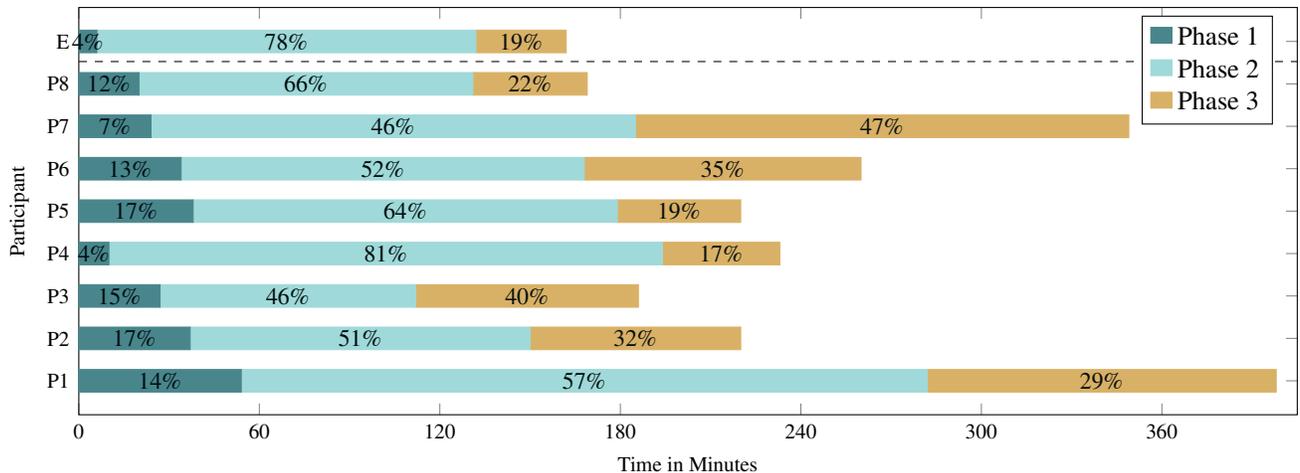

Figure~\ref{fig:solutiontime} shows the time each participant required for solving the Study Task, together with a break down of the times they spent on each of the three phases. On average, participants spent $254$~minutes ($SD=74$~minutes) on task. We also calculated the group average of relative time spent on each of the three phases, by averaging the relative times of all eight participants: On average, they spent $12\%\ (SD=4\%)$ of their time in Phase~1, Phase~2 consumes $58\%\ (SD=11\%)$, and Phase~3 took $30\%\ (SD=10\%)$ of participants' time solving the task.

\subsubsection{Executed Scripts per Phase}
On average, participants executed $117$~scripts ($SD=32$) while solving the Study Task. We calculated the relative number of executed scripts per phase by averaging relative number of scripts per participant: $15\%\ (SD=9\%)$ of scripts were executed in Phase~1, $55\%\ (SD=13\%)$ in Phase~2, and $30\%\ (SD=12\%)$ in Phase~3. The number of executed scripts per phase is represented in Appendix~\ref{appendix:results}, Table~\ref{tab:scripts}.

\subsubsection{Manual Component Inspections per Phase}
On average, participants manually inspected $123$ ($SD=82$) netlist components while working on the Study Task. We calculated relative numbers of inspected components per phase by averaging relative numbers per participant. Participants conducted $48\%\ (SD=35\%)$ of their manual inspections in Phase~1,  $42\%\ (SD=38\%)$ in Phase~2, and $10\%\ (SD=8\%)$ in Phase~3. Table~\ref{tab:gui} in Appendix~\ref{appendix:results} shows detailed statistics on manual component inspections for all participants.

\subsection{\ac{HRE} Strategies (RQ2)}
\label{casestud}
In this section, we perform an in-depth exploration of \ac{HRE} strategies per phase for Participants~1 (P1) and 8 (P8), the least and most efficient overall participants with respect to solution time. While P1 required a total time of 398~minutes, P8 solved the task in 169~minutes. Both participants spent a similar amount of relative time in Phases~1 to 3 (cf. Figure~\ref{fig:solutiontime}), and had comparable total amounts of manual netlist interactions (P1: $127$, P8: $90$). P8 executed the least (62) and P1 the third-most (132) scripts. We want to illustrate that this was not a problem of poor programming skills: Whereas only $47\%$ of the scripts written by P8 were syntactically correct, P1 executed $73\%$ syntactically correct scripts.
A detailed overview of the assigned progress scores per phase as introduced in Section~\ref{subsubsec:behavanalysis} is shown in Appendix~\ref{appendix:results}, Table~\ref{tab:casestud}.
In the case of executed scripts, a progress score of 0 --- which indicates stagnation --- dominates all phases but Phase~1 of P8. Manual analyses have crucial impact (score of~3) on progress in Phase~1 for both participants, and on Phase~2 for P8.
Overall, script-based progress dominates progress made through manual analyses. A visualization of participants' P1 and P8 progress over time is shown in Figure~\ref{fig:strategy2} in Appendix~\ref{appendix:results}. The figure visually complements the exploration of strategies applied by P1 and P8 in Phases~1, 2 and~3 as presented below.

\subsubsection{Phase~1 (Candidate Identification)}
P1 required 54~minutes to solve Phase~1, while P8 identified the correct candidates in 20~minutes. Both participants started with manual analysis and were able to achieve initial progress by selecting relevant components: P1 detected the first relevant components after 7~minutes, and P8 found the first relevant components after 4~minutes. 
After initial progress had been made, P8 implemented his concept of structural candidate identification in the very first script (minute~18), which then only needed minor adjustments in order to finish Phase~1.
P1 similarly implemented his crucial idea for Phase~1 in the first script (minute~26) but needed several script iterations in order to successfully identify candidates.
Despite their similar solution strategies with respect to interactions of manual and script-based analyses, our manual script review revealed that P1 chose a less efficient and more complex implementation to detect the candidates as indicated by a larger search space and a deeper nesting of the algorithm.

\subsubsection{Phase~2 (Candidate Verification)}
Participants~1 and 8 spent 111 respectively 228~minutes in order to functionally verify the candidates.
P8 made immediate progress through manual inspection of relevant components. Afterwards, P8 progressed towards the verification of candidates via the development of scripts and single manual inspections without significant periods of stagnation. Only between minutes~89 and 99, the participant could not advance the problem of data extraction from netlist components via script.
In our manual reviews of scripts, we observed that P8 applied a divide-and-conquer strategy, and considered the overall objective --- removing the watermark signatures --- already in his approach of extracting them in Phase~2.

On the contrary, P1 started with an initial stagnation period (minute~54 to 78), where three scripts were executed without noticeable direction. His initial progress in Phase~2 was sparked through the manual inspection of relevant components, although there were still irrelevant components among the inspected. The next significant stagnation period (minute~131 to 203) concerned the same problem causing the 10-minute stagnation of P8. Another period of stagnation lasted from minute~210 to 242, where P1 inspected irrelevant components.
Manual review of scripts revealed that P1 developed an efficient but complex algorithm for Phase~2 including bitwise binary processing depending on several conditions.

\subsubsection{Phase~3 (Realization)}
The realization of the underlying objective took 116~minutes for P1 and 37~minutes in the case of P8.
Consequently, P8 moved quickly towards the solution without manual netlist inspections. In the stagnation period from minute~153 to 168, he merely tried to fix printing methods.
P1 had a significant stagnation period from minute~305 to 352, where the manipulation of binary strings represented a significant hurdle.
After this stagnation period, manual component inspections served as impulse for further progress. In the last stagnation period, P1 went back to the algorithm from Phase~1 to replace hard-coded data with variables.

\subsection{Cognitive Factors and \ac{HRE} (RQ3)}
\label{cognitiveanalysis}
The small sample size rendered statistical analyses of the influences of cognitive factors on \ac{HRE} performance impossible. 
We do, however, want to briefly describe one interesting finding which can be discussed as an incentive for future research on cognitive factors in \ac{HRE}. Our data suggest a potential negative correlation between \acf{WM} scores and the overall solution time of the Study Task. The descriptive data shows that participants with higher scores in \ac{WM} tend to solve the task quicker than participants with lower \ac{WM} scores. P8 had the highest \ac{WM} score (126) and achieved the shortest time on task with 169~minutes. Meanwhile, the least efficient participant (P1) with a time on task of 398~minutes had a lower \ac{WM} score of~108. There appears to be one outlier in the data (P6) which will be discussed in Section~\ref{disc:cogfac}. The descriptive and visual analyses by scatter plots for both cognitive factors \acf{PS} and \acf{PR} did not show comparable results. Appendix~\ref{appendix:results}, Table~\ref{tab3:cognition} summarizes the descriptive data of cognitive factors and solution time.

\begin{figure}[htbp]
	\begin{tikzpicture}
	\begin{axis}[
	ylabel=Working Memory Score,
	xlabel=Total Solution Time in Minutes,
	scaled x ticks = false,
	height=6cm,
	width=\linewidth,
	ymin    = 80,
	ymax    = 140,
	xmin    = 8000,
	xmax    = 26000,
	ytick   = {80, 90, ..., 140},	
	xtick   = {8000, 9000, 12000, 15000, 18000, 21000, 24000},
	xticklabels= {, 150, 200, 250, 300, 350, 400}]
	\addplot[color=blue, only marks,mark=*, mark size=1.5pt] coordinates {(23851, 108)} node[above right=0.05cm, color=black, draw, inner sep=0.4mm]{\footnotesize$P1$} ;
	\addplot[color=blue, only marks,mark=*, mark size=1.5pt] coordinates {(13088, 126)} node[above right=0.05cm, color=black, draw, inner sep=0.4mm]{\footnotesize$P2$} ;
	\addplot[color=blue, only marks,mark=*, mark size=1.5pt] coordinates {(13921, 115)} node[below right=0.05cm, color=black, draw, inner sep=0.4mm]{\footnotesize$P4$} ;
	\addplot[color=blue, only marks,mark=*, mark size=1.5pt] coordinates {(13225, 118)} node[above right=0.05cm, color=black, draw, inner sep=0.4mm]{\footnotesize$P5$} ;
	\addplot[color=blue, only marks,mark=*, mark size=1.5pt] coordinates {(15631, 92)} node[below right=0.05cm, color=black, draw, inner sep=0.4mm]{\footnotesize$P6$} ;
	\addplot[color=blue, only marks,mark=*, mark size=1.5pt] coordinates {(20906, 112)} node[above right=0.05cm, color=black, draw, inner sep=0.4mm]{\footnotesize$P7$} ;
	\addplot[color=blue, only marks,mark=*, mark size=1.5pt] coordinates {(10541, 126)} node[below right=0.05cm, color=black, draw, inner sep=0.4mm]{\footnotesize$P8$} ;		
	
	\addplot [domain=5000:26000, samples=100] {133 - 0.00119*x};
	\end{axis}
	\end{tikzpicture}
	\label{fig:scatterplot}
	\caption{Scatter plot of solution time ($x$-axis) and \acl{WM} scores ($y$-axis) with trend line. P3 did not participate in the cognitive tests.}
\end{figure}
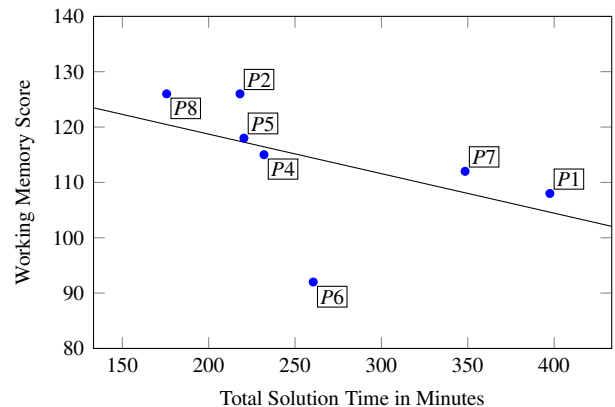
\section{Discussion}
\label{disc}
In this section, we discuss implications of the results presented in Section~\ref{sec_Results} with respect to our four research questions.
\subsection{\ac{HRE} Phase Model (RQ1)}
\label{subsec:disc:rq1}
We were able to detect the proposed three-phase model for netlist reverse engineering for all participants of our study (cf. Figure~\ref{fig:solutiontime}).
Thus, we suggest that passing through all phases in their respective order is essential to solve the underlying task.
Our data revealed differences in absolute and relative times spent per phase, implicating their respective levels of difficulty.
Moreover, we observed \ac{HRE} as an interwoven process of manual and script-based netlist interactions. 
The observed proportions of manual component inspections per phase imply a high relevance of visual inspections for Phases~1 and 2. Even though amounts of manual interactions varied between participants as indicated by high standard deviations, we could observe purposeful usage for every participant.
In the following, we briefly discuss our results with respect to each phase.

\subsubsection{Phase~1 (Candidate Identification)}
The relatively short solution times observed in this phase indicate that the identification of candidates, that is, relevant components, could be solved efficiently via structural analyses as described in Section~\ref{phasemod}. Since most manual netlist interactions were conducted in Phase~1, we deduce that they play an important role in order to detect and inspect starting points (e.g., global in- and outputs, or nets with a great number of sinks) for reverse engineering.
The identification of candidates in Phase~1 is a necessary precondition to conduct targeted candidate verification in Phase~2. It is therefore --- despite its relatively short duration as observed in this exploratory study --- crucial for the overall \ac{HRE} process.

\subsubsection{Phase~2 (Candidate Verification)}
The significant portion of total solution time spent on Phase~2 implies that candidate verification was the most challenging subtask for our participants. In order to verify candidates, participants developed algorithms which functionally analyzed single candidates. We assume that the development of such an algorithm is a challenging problem. The results suggest that manual analyses in Phase~2 supported most participants in verifying candidates and developing their algorithms.
A sound solution of Phase~2 is very important 
as preparation for the specific \ac{HRE} objectives realized in Phase~3. 

\subsubsection{Phase~3 (Realization)}
In Phase~3, reverse engineers conduct goal-oriented actions on previously identified targets by means such as interpretation, manipulation, simulation, or a combination thereof.
In our case, watermarks had to be removed from the netlist via manipulation. Although netlist manipulation was an essential part of the training tasks, it was difficult to transfer those skills to the application of removing watermarks as indicated by a portion of $30\%$ of total solution time.
Participants applied manual interactions only sparsely in Phase~3, suggesting that they have less impact here compared to Phases~1 and 2.



\subsection{\ac{HRE} Strategies (RQ2)}
\label{subsec:disc:rq2}
In this section we discuss our results from the in-depth exploration of \ac{HRE} strategies of the most and least efficient participants (cf. Section~\ref{casestud}).

The comparable amounts of manual netlist interactions for P1 and P8 indicate their importance for \ac{HRE} regardless of the reverse engineers' efficiency. Different amounts of executed scripts between both participants are an evidence that the faster participant made more progress per script. However, our data suggests that P8 was not the better programmer than P1, implying that other factors contribute to a higher level of progress per script.

With respect to our progress score assignments (cf. Table~\ref{tab:casestud}), we note that scores of 0 indicating stagnation dominate the overall \ac{HRE} progress. This supports the assumed high complexity of netlist reverse engineering even for medium-sized netlists as in the underlying Study Task. The observation that script-based progress dominates progress made by manual analyses implies the nature of manual analysis methods as supporting factor for \ac{HRE}.

In the following passages, we discuss the progress over time of P1 and P8 with special emphasis on periods of stagnation (cf. Figure~\ref{fig:strategy2} in Appendix~\ref{appendix:results}).

\subsubsection{Phase~1 (Candidate Identification)}
In Phase~1, both participants followed similar strategies based on structural netlist exploration via semi-automated scripts. 
The fact that initial progress was made through manual component inspections by both participants further supports the indication from Section~\ref{subsec:disc:rq1} that manual inspections play an important role at early stages of netlist reverse engineering.
Longer duration of Phase~1 for P1 was caused by a more complex algorithmic approach for candidate identification which is harder to implement.

\subsubsection{Phase~2 (Candidate Verification)}
Both participants followed different strategies in Phase~2 in order to verify candidates identified in Phase~1.
P1 started with manual inspection of several relevant components, which then were manually divided into subgroups of candidates.
Subsequently, P8 started --- supported by further manual analysis --- the development of an algorithm applying a divide-and-conquer strategy. Thus, we assume that manual analysis is a crucial factor for choosing this strategy and therefore to solve Phase~2 in a shorter amount of time. Towards the end of Phase~2, P8 applied his algorithm in a slightly adapted version to the other subproblems. Consequently, the divide-and-conquer strategy seems to be an efficient approach for solving \ac{HRE} problems in this phase. 

On the contrary, no clear strategy is identifiable for P1 initially. We observed that P1 tried to continue developing the candidate extraction algorithm from Phase~1 for this phase. We assume that P1 did not develop such a concrete plan as P8 based on the long-lasting stagnation periods observed throughout Phase~2. 

Additionally, we monitored a significant stagnation period of 70~minutes, where P1 tried to read-out data from candidates; P8 solved the same problem within 10~minutes. We assume that P8 was able to apply prior knowledge from training tasks faster, since the they taught all necessary methods to solve this problem. A further stagnation period during a manual component selection could be observed for P1. We hypothesize that P1 tried to explore new points of interest in order to continue the verification of his candidates. This is a further hint that P1 applied a sub-optimum strategy.

\subsubsection{Phase~3 (Realization)}
Forward-thinking of P8 enabled the fast solution time in Phase~3 because the participant could adapt the existing algorithms from Phase~2 in order to remove the watermark. In contrast, P1 struggled as indicated by a long stagnation period in which the participant tried to remove the watermarking on a binary level.
Additionally, P1 still needed visual input by manual netlist interactions in order to execute the planned strategy, which is a time-consuming process. 
In the last stagnation period, P1 tried to improve the algorithm for candidate identification originally implemented in Phase~1.

\subsection{Cognitive Factors and \ac{HRE} (RQ3)}
\label{disc:cogfac}
In the light of RQ3, the conduction of statistical analyses was infeasible based on the small sample size. Nevertheless, we reported a promising result concerning a possible negative correlation between scores in \acf{WM} and solution time. The term \ac{WM} is defined as a brain system which enables the storage of sensory input (e.g., visual input) in immediate awareness and the manipulation of that input in order to solve complex cognitive tasks like problem solving \cite{baddeley1992working}. It consists of several parts, for example, the central executive (attentional-controlling system) \cite{2012:baddeley}. Based on a descriptive analysis the data suggested that participants with higher \ac{WM} scores tend to solve the task faster than participants with lower \ac{WM} scores. Due to missing statistical analyses, we can only hypothesize that the \ac{WM} might play a role in solving \ac{HRE} problem tasks and future studies should statistically investigate the interplay between \ac{WM} and the efficiency of solving \ac{HRE} tasks. 

Nevertheless, we try to explain differences in problem solving strategies of the participants by referring to the functionalities of the \ac{WM}. Baddeley and Hitch postulated that the \ac{WM} is central for solving cognitively difficult tasks like problem solving by performing mental operations on temporally stored information~\cite{baddeley1974working}. Against this background, it might be possible that P8 with the highest \ac{WM} score achieved a more efficient solution than P1 who had a lower \ac{WM} score. P8's strong \ac{WM} might have supported the participant in mentally dividing the main problem into several sub problems, and helped in selecting further actions in order to achieve sub goals. Moreover, we assume that P8's stronger \ac{WM} might have helped to conclude a stagnation period in Phase~2 more efficiently (10~minutes) by quicker activating relevant prior knowledge compared to P1 who needed over 70~minutes. Both participants struggled with the same problem in reading out relevant information from candidates. Although both participants acquired the same amount of \ac{HRE} knowledge and \ac{HRE} skills (e.g., methods to read out data from the identified candidates) during the training phase, P8 was quicker in solving that problem. 
We hypothesize, that the stronger \ac{WM} enabled P8 to activate stored information from the long-term memory in order to analyze and work with inputs the participant stored in immediate awareness in the \ac{WM}. 
This ability of activating prior knowledge is described by the term chunks~\cite{chase1973perception}, which postulates that the working load of the \ac{WM} can be reduced by activation of knowledge structures. Those chunks can boost the immediate memory of the \ac{WM} and are connected to the level of expertise. A stronger \ac{WM} could have been advantageous in producing a faster and more efficient solution.

Moreover, the scatter plot also revealed an outlier (P6) with the lowest \ac{WM} score. A possible explanation for this outlier is that the measurements of \ac{WM} abilities could have been influenced by uncontrolled variables. Prior work showed that the performances in \ac{WM} tasks can be impaired by several variables, such as chronic psychological stress~\cite{mizoguchi2000chronic}, acute psychological stress~\cite{qin2009acute}, negative emotions like anxiety~\cite{moran2016anxiety}, or neural disorders~\cite{willcutt2005validity}.



\subsection{Hypotheses for Cognitive Obfuscation (RQ4)}
Hardware obfuscation is understood as design method that impedes reverse engineering. Even though it should not be considered a silver bullet for hardware security, it can be a useful tool, for example, for improving \ac{IP} protection or increasing the cost factor of an attack. Two important observations of our exploratory study are that (i) all participants progressed through a unique phase model and that (ii) the principal methods with respect to manual and script-based analyses employed by the participants were similar. This allows us to derive hypotheses for hardware obfuscation that take this process steps into account.

%

\subsubsection{Obfuscation delaying \ac{HRE} Phases}
Our data showed that Phase~1, the necessary first step of netlist reverse engineering, can be solved efficiently. 
In order to increase time on Phase~1, we suggest the following approach for cognitive obfuscation: The netlist should be composed of many equally-looking regular structures, 
for example, by employing dummy wires or camouflaged gates \cite{vijayakumar2016physical}. Only few of those structures should contain the actual \ac{HRE} targets. Thus, Phase~1 search techniques will return a considerable number of candidates. As a result Phase~1 will be slowed down through the amount of candidates found, as well as Phase~2 since all candidates have to be verified functionally.

\subsubsection{Obfuscation impeding \ac{HRE} Strategies}
Hardware reverse engineers have to develop strategies that are tailored to their specific objective. From a problem solving perspective, Dörner and Funke concluded that a universal problem solving strategy which can be applied to solve different problems does not exist~\cite{dorner2017complex}. Against this background, it seems promising to apply different cognitive obfuscation methods for a given netlist which are similar in appearance but require individual solution strategies. This will force the reverse engineer to develop and apply various strategies rather than using a universal one. 



\subsubsection{Obfuscation against Cognitive Abilities}
Based on our initial findings on the role of \acf{WM}, it is possible that the performance in \ac{HRE} might be correlated to levels of \ac{WM}. If future studies can support this hypothesis with statistical data, the development of countermeasures could aim to overload cognitive capabilities of a reverse engineer. One important aspect in this context is the capacity of the \ac{WM}. Even though the commonly assumed limit of seven items \cite{miller1956magical} can be extended by activating knowledge structures (chunks) stored in the long-term memory \cite{ericsson1995long, gobet1996templates, chase1973perception}, the capacity of items that can be stored will always be a rather moderate number. 
We hypothesize that cognitive obfuscation could overload the capacity of the \ac{WM}. A possible approach might be the development of structures which force the analyst to apply a combination of many different strategies simultaneously (e.g., simulation in Phase~3 in combination with structural and functional analyses in Phases~1 and 2) which might overload the capacity of the \ac{WM}. 

\subsection{Limitations and Future Work}
\label{limitations}
While our exploratory study provides important initial insights into the technical and cognitive processes underlying \ac{HRE}, there are certainly limitations. First, this research is limited by the small sample size which did not allow us to conduct any statistical analyses. With respect to cognitive factors, future research with larger sample sizes could quantify the influences of \ac{WM} on \ac{HRE} processes more accurately. In this context, it would also be interesting to explore to what extent prior knowledge and chunks are activated during \ac{HRE}.

Second, our results are based on the behavioral analyses of cyber security students who acquired relevant \ac{HRE} skills and knowledge during an extensive training phase. Even though we worked with students, we took precautions to compare our sample and the Study Task by recruiting an \ac{HRE} expert. The descriptive analyses of this expert served as a sanity check for 
our findings, the approximation of experienced revers engineers with our student sample, and the appropriate difficulty of the Study Task.
Our results, discussion and takeaways regarding strategies were obtained analyzing two participants. Further detailed analyses of additional participants’ and of the expert’s \ac{HRE} problem solving strategies would have been worthwhile.
However, given that the primary goal was to provide first insights into the technical and cognitive processes of \ac{HRE}, we feel that our analysis is warranted and led to interesting results worth reporting.
We hope that our exploratory study will trigger follow-up work which focuses on the in-depth analysis of problem solving strategies in \ac{HRE}.  
In the future, qualitative data such as interview data may complement the comprehension of technical and cognitive processes in \ac{HRE} and could potentially lead to a more nuanced view of the strategies the participants followed.


Third,  we developed our \ac{HRE} model based on the behavioral analyses of reverse engineers for a single, medium-complex \ac{HRE} task and it remains unclear whether the phases transfer to different \ac{HRE} settings. Nevertheless, it seems plausible that the three phases can be found in other \ac{HRE} tasks too, but more complex real-world tasks could potentially lead to an extension of the phase model, for example, by discovering further sub phases that are universally observed.

Finally, we proposed first hypotheses for cognitive obfuscation techniques. Future work should evaluate which cognitively difficult tasks are suitable to raise time on \ac{HRE} tasks and how they can be implemented in netlists.
\section{Conclusion}
The motivation behind this work is to take a first step towards a better understanding of technical and cognitive processes in \ac{HRE}, an area with little prior published work despite its importance in commercial and national security contexts.
We conducted an exploratory behavioral study with eight participants who solved a realistic \ac{HRE} task.
We examined the approximation of the participants as experienced reverse engineers and the appropriate difficulty of the Study Task through behavioral analyses of an \ac{HRE} expert.
We found that the three-phase model for \ac{HRE}, which we had postulated, was in fact executed by all participants (as well as by the expert). 
The analysis of behavioral data showed that the two principal types of actions, manual and automatic analyses, are closely interwoven during a given \ac{HRE} task. 
The analysis of cognitive prerequisites of the participants indicates that the \acf{WM} might play an important role in solving \ac{HRE} problems.

Our study suggests several new research directions that can be worthwhile to explore. 
It seems promising to extend the study to a larger population of participants and, in particular, to observe true \ac{HRE} experts using a similar methodology as presented here.
Another especially interesting research direction lies in the design of a novel class of countermeasures against \ac{HRE}, which will take cognitive limitations of reverse engineers into account.

\section*{Acknowledgements}
We would like to thank Yasemin Acar, Maximillian Golla, as well as our anonymous shepherd and all reviewers for their constructive feedback on this work.
We are also grateful to Oliver Gebhard and René Walendy for supporting us with data analysis. 
Finally, we thank Carl for contributing the \enquote{Carlgorithm} at an early stage of this research.

This work was supported in part by DFG Excellence Strategy grant 39078197 (EXC 2092, CASA), through ERC grant 695022 and NSF grant CNS-1563829.

\bibliographystyle{plain}
\bibliography{bibliography}

\appendix
\section*{Appendix}
\section{Survey Instruments}
\label{appendix:questionnaires}
\subsection*{Participant Questionnaire}
\begin{compactenum}
	\item Please enter your pseudonym: \_\_\_\_\_\_\_\_\_\_
	\item How old are you? Please indicate your age in years. \_\_\_\_\_\_\_\_\_\_
	\item What is your target degree? Please select one answer.
	\begin{compactitem}
		\item[$\radiobutton$] Bachelor of Science
		\item[$\radiobutton$] Master of Science
		\item[$\radiobutton$] Other: \_\_\_\_\_\_\_\_\_\_
		\item[$\radiobutton$] Prefer not to answer
	\end{compactitem}
	\item What is your major? Please select one answer.
		\begin{compactitem}
		\item[$\radiobutton$] Cyber Security
		\item[$\radiobutton$] Electrical Engineering
		\item[$\radiobutton$] Computer Science
		\item[$\radiobutton$] Other: \_\_\_\_\_\_\_\_\_\_
		\item[$\radiobutton$] Prefer not to answer
	\end{compactitem}
	\item Please indicate the number of semesters you have studied so far. \_\_\_\_\_\_\_\_\_\_
\end{compactenum}
\subsection*{Expert Questionnaire}
\begin{compactenum}
	\item How old are you? Please indicate your age in years: \_\_\_\_\_\_\_\_\_\_
	\item What is your highest level of education? Please select one answer.
		\begin{compactitem}
		\item[$\radiobutton$] Bachelor of Science
		\item[$\radiobutton$] Master of Science
		\item[$\radiobutton$] Bachelor of Arts
		\item[$\radiobutton$] Master of Arts
		\item[$\radiobutton$] Ph.D.
		\item[$\radiobutton$] Other: \_\_\_\_\_\_\_\_\_\_
	\end{compactitem}
	\item What is your current job position? What are your responsibilities? \_\_\_\_\_\_\_\_\_\_
	\item On a scale 1 to 5, how would you asses your hardware reverse engineering skill level? Please indicate your skill level on the 5-point scale from 1 (being a beginner) and 5 (being an expert) by choosing one answer.\\
	$\radiobutton$ 1 (Beginner) \hspace{.25cm} $\radiobutton$ 2 \hspace{.25cm} $\radiobutton$ 3 \hspace{.25cm} $\radiobutton$ 4 \hspace{.25cm} $\radiobutton$ 5 (Expert)
	\item How many total years of experience do you have with hardware reverse engineering? Please indicate your answer in years. \_\_\_\_\_\_\_\_\_\_
\end{compactenum}
\section{Full Results}
\label{appendix:results}
\vspace*{-1cm}
\begin{table}[htb]
	\small
	\begin{center}
		\hspace*{-.25cm}
		\caption{Amount of excluded inactivity periods over 60 minutes per participant and phase.}
		\label{tab:inacticity}
		\resizebox{\linewidth}{!}{
			\begin{tabular}{lccccccccc}
				\toprule
				&  \textbf{P1}  & P2  &  P3  &  P4  &  P5  & P6  &  P7  & \textbf{P8} & \textbf{E}  \\ \cmidrule(lr){1-9} \cmidrule(lr){10-10}
				Phase~1 & 0  & 1 & 0  &  0  & 0  & 0 &  0  & 0 & 0 \\ \cmidrule(lr){2-9} \cmidrule(lr){10-10}
				Phase~2 & 2  & 0 & 0  & 2   & 1 & 0 & 1  & 0 & 1 \\ \cmidrule(lr){2-9} \cmidrule(lr){10-10}
				Phase~3 & 3  & 0 & 3  & 0  & 0  & 1 & 1  & 1 & 0 \\ \cmidrule(lr){1-9} \cmidrule(lr){10-10}
				Total   & \textbf{5} & 1 & 3 & 2 & 1 & 1 & 2 & \textbf{1} & \textbf{1} \\
				\bottomrule
			\end{tabular}
		}
	\end{center}
\end{table}
\vspace*{-1.2cm}
\begin{table}[H]
	\small
	\begin{center}
		\hspace*{-.25cm}
		\caption{Amount of executed scripts per participant and phase. For comparison, the experts' amount of executed scripts is represented in the rightmost column.}
		\label{tab:scripts}
		\resizebox{\linewidth}{!}{
		\begin{tabular}{lccccccccc}
			\toprule
			&  \textbf{P1}  & P2  &  P3  &  P4  &  P5  & P6  &  P7  & \textbf{P8} & \textbf{E}  \\ \cmidrule(lr){1-9} \cmidrule(lr){10-10}
			Phase~1 & 28  & 21 & 33  &  3  & 42  & 20 &  4  & 3 & 5 \\ \cmidrule(lr){2-9} \cmidrule(lr){10-10}
			Phase~2 & 60  & 51 & 72  & 111 & 103 & 35 & 49  & 39 & 122 \\ \cmidrule(lr){2-9} \cmidrule(lr){10-10}
			Phase~3 & 44  & 27 & 21  & 23  & 30  & 40 & 57  & 16 & 37 \\ \cmidrule(lr){1-9} \cmidrule(lr){10-10}
			Total   & \textbf{132} & 99 & 126 & 137 & 175 & 95 & 110 & \textbf{62} & \textbf{164} \\
			\bottomrule
		\end{tabular}
	}
	\end{center}
\end{table}
\begin{table}[htbp]
	\small
	\begin{center}
		\hspace*{-.25cm}
		\caption{Amount of manual component inspections per participant and phase. For comparison, the experts' amount of manual inspections is shown in the rightmost column.}
		\label{tab:gui}
		\resizebox{\linewidth}{!}{
		\begin{tabular}{lccccccccc}
			\toprule
			&  \textbf{P1}  & P2  & P3  &  P4  &  P5  & P6  &  P7  & \textbf{P8}  &  \textbf{E}  \\\cmidrule(lr){1-9} \cmidrule(lr){10-10}
			Phase~1 & 93  & 66 & 0  &  4  & 217 & 16 & 47  & 23  & 8  \\\cmidrule(lr){2-9} \cmidrule(lr){10-10}
			Phase~2 & 19  & 4  & 37 & 162 & 8   & 0  & 93  & 67 & 194 \\\cmidrule(lr){2-9} \cmidrule(lr){10-10}
			Phase~3 & 15  & 0  & 1  & 21  & 59  & 4  & 24  & 0  & 2  \\\cmidrule(lr){1-9} \cmidrule(lr){10-10}
			Total   & \textbf{127} & 70 & 38 & 187 & 284 & 20 & 164 & \textbf{90} & \textbf{204} \\ \bottomrule
		\end{tabular}
	}
	\end{center}
\end{table}
\begin{table}[htbp]
	\small
	\begin{center}
	    \hspace*{-.25cm}
		\caption{Overview of assigned progress scores for executed scripts and manual analysis assigned to P1, P8, and the expert as explained in Section~\ref{subsubsec:behavanalysis}. Dashes indicate zero occurrences.}
		\label{tab:casestud}
		\resizebox{\linewidth}{!}{
		\begin{tabular}{llcccccccccccc}
			\toprule
			& & \multicolumn{4}{c}{P1} & \multicolumn{4}{c}{P8} &  \multicolumn{4}{c}{Expert} \\ \cmidrule(lr){3-6} \cmidrule(lr){7-10} \cmidrule(lr){11-14}
			& Score & 0 & 1 & 2 & 3 & 0 & 1 & 2 & 3 & 0 & 1 & 2 & 3\\ \cmidrule[1pt](lr){1-14}
			\multirow{2}{*}{Phase~1} & Script & 22 & 3 & 3 & 1 & 1 & - & 1 & 1 & 1 & 2 & 1 & 1\\ \cmidrule(lr){3-6} \cmidrule(lr){7-10} \cmidrule(lr){11-14}
			& Manual & 1 & - & - & 1 & 1 & - & - & 1 & 8 & - & - & -\\ \cmidrule[1pt](lr){1-14}
			\multirow{2}{*}{Phase~2} & Script & 48 & 7 & 4 & - & 29 & 5 & 2 & 2 & 92 & 26 & 5 & 1\\ \cmidrule(lr){3-6} \cmidrule(lr){7-10} \cmidrule(lr){11-14}
			& Manual & 7 & 2 & - & - & 1 & 3 & - & 1 & 5 & 6 & 2 & 1\\ \cmidrule[1pt](lr){1-14}
			\multirow{2}{*}{Phase~3} & Script & 36 & 5 & 2 & - & 12 & 3 & 1 & - &25 & 7 & 4 & 1 \\ \cmidrule(lr){3-6} \cmidrule(lr){7-10} \cmidrule(lr){11-14}
			& Manual & 2 & 3 & - & - & - & - & - & - & - & 1 & - & -\\ 
			\bottomrule
		\end{tabular}
	}
	\end{center}
\end{table}
\begin{table}[htb]
	\small
	\begin{center}
	\hspace*{-.25cm}
	\caption{Scores of the cognitive factors \acf{WM}, \acf{PS} and \acf{PR} per participant and time spent on the Study Task in minutes (P3 did not participate in the cognitive tests).}
	\label{tab3:cognition}
	\resizebox{\linewidth}{!}{
		\begin{tabular}{lcccccccc}
			\toprule
			&  P1  &  P2  & P3 &  P4  &  P5  &  P6  &  P7  &  P8  \\ \cmidrule(lr){1-9}
			\ac{WM} & \textbf{108} & 126 &   & 115 & 118 & 92  & 112 & \textbf{126} \\ \cmidrule(lr){2-9}
			\ac{PS} & 106 & 146 &   & 146 & 119 & 109 & 119 & 117 \\ \cmidrule(lr){2-9}
			\ac{PR} & 104 & 129 &   & 100 & 115 & 100 & 104 & 106 \\ \cmidrule(lr){2-9}
			Time & \textbf{398} & 218 & 186  & 232 & 220 & 261 & 348 & \textbf{176} \\
			\bottomrule
		\end{tabular}
	}
	\end{center}
\end{table}
\newpage

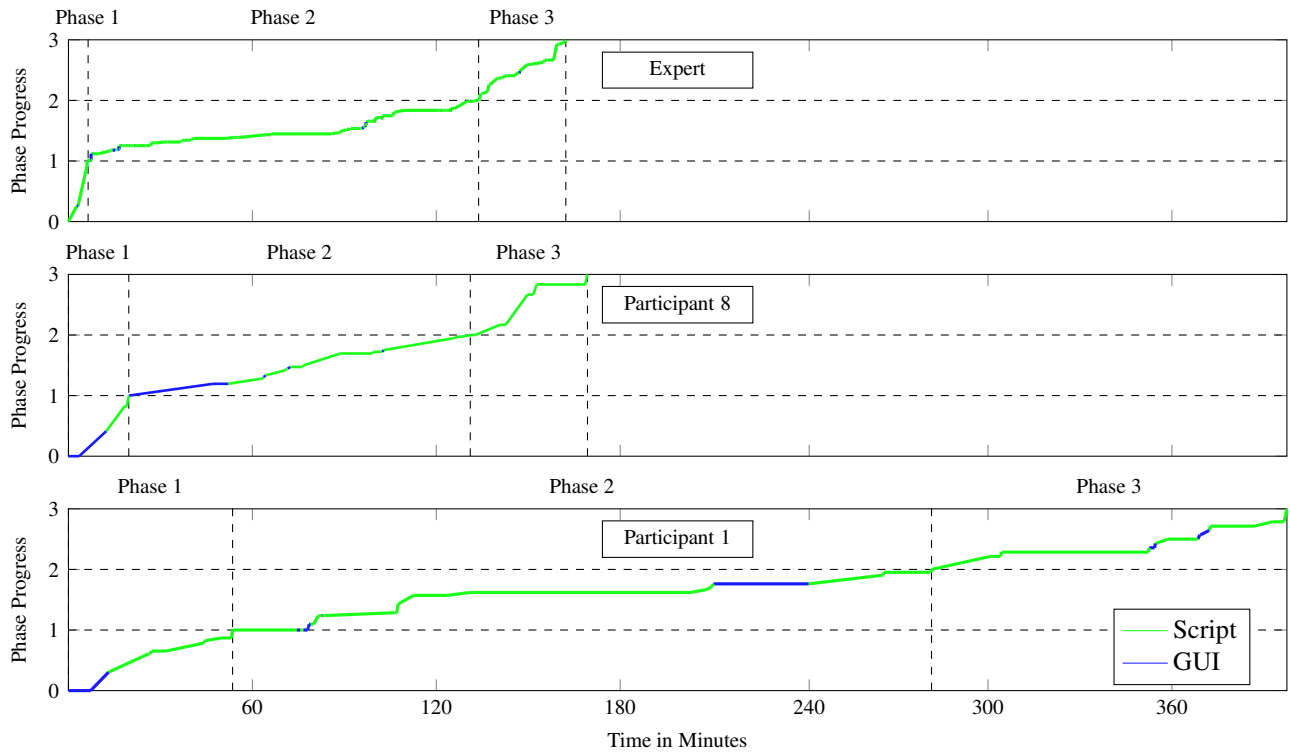
\begin{figure*}
	\centering
	\begin{tikzpicture}
	\begin{groupplot}[
	group style={
		group name=my plots,
		group size=1 by 3,
		xlabels at=edge bottom,
		xticklabels at=edge bottom,
		vertical sep=20pt
	},
	width=\textwidth,
	height=4cm,
	xmin    = 0,
	xmax    = 23851,
	ymin    = 0,
	ymax    = 3,
	xtick   = {0, 3600, 7200, 10800, 14500, 18000, 21600, 25200},
	xticklabels= { , 60, 120, 180, 240, 300, 360},
	ytick   = {0, 1, 2, 3},
	yticklabels={0, 1, 2, 3},
	xlabel  = {Time in Minutes},
	scaled x ticks = false,
	colormap={test}{color=(c5) color=(c4)}
	]

	\nextgroupplot[clip=false, ylabel=Phase Progress]
	
	\addlegendimage{no markers,red}
	\addlegendimage{no markers,blue}

	
	\addplot[mark=none, black, dashed] coordinates {(0,1) (23851,1)};
	\addplot[mark=none, black, dashed] coordinates {(0,2) (23851,2)};
	
	\addplot[mark=none, black, dashed] coordinates {(385,0) (385,3.1)};
	\addplot[mark=none] coordinates {(385,3.1)} node [above] {\footnotesize Phase 1};
	
	\addplot[mark=none, black, dashed] coordinates {(8029,0) (8029,3.1)};
	\addplot[mark=none] coordinates {(4207,3.1)} node [above] {\footnotesize Phase 2};
	
	\addplot[mark=none, black, dashed] coordinates {(9736,0) (9736,3.1)};
	\addplot[mark=none] coordinates {(8882,3.1)} node [above] {\footnotesize Phase 3};
	
	\addplot[mark=none] coordinates {(11925,2.5)} node [midway,draw,rectangle, minimum width=2cm]  {\footnotesize Expert};
	

	
	\addplot [
	mesh,
	line width=1.2pt,
	scatter, mark=none,
	scatter src=explicit symbolic,
	shader=flat corner,
	point meta=\thisrow{color}
	] table [col sep=semicolon] {figures/graphs/Expert1.csv};

	\nextgroupplot[clip=false, ylabel=Phase Progress]

	
	\addplot[mark=none, black, dashed] coordinates {(0,1) (23851,1)};
	\addplot[mark=none, black, dashed] coordinates {(0,2) (23851,2)};

	\addplot[mark=none, black, dashed] coordinates {(0,0) (0,3.1)};
	\addplot[mark=none] coordinates {(590,3.1)} node [above] {\footnotesize Phase 1};
	
	\addplot[mark=none, black, dashed] coordinates {(1181,0) (1181,3.1)};
	\addplot[mark=none] coordinates {(4523,3.1)} node [above] {\footnotesize Phase 2};
	
	\addplot[mark=none, black, dashed] coordinates {(7866,0) (7866,3.1)};
	\addplot[mark=none] coordinates {(9013,3.1)} node [above] {\footnotesize Phase 3};
	
	\addplot[mark=none, black, dashed] coordinates {(10160,0) (10160,3.1)};
	
	\addplot[mark=none] coordinates {(11925,2.5)} node [midway,draw,rectangle, minimum width=2cm]  {\footnotesize Participant 8};


	\addplot [
	mesh,
	line width=1pt,
	scatter, mark=none,
	scatter src=explicit symbolic,
	shader=flat corner,
	point meta=\thisrow{color},
	] table [col sep=semicolon] {figures/graphs/Twinrova.csv};

	\nextgroupplot[clip=false, legend entries={Script,
		GUI},legend cell align={left},
	legend pos=south east,ylabel=Phase Progress]
	
	\addlegendimage{no markers,c4}
	\addlegendimage{no markers,c5}

	
	\addplot[mark=none, black, dashed] coordinates {(0,1) (23851,1)};
	\addplot[mark=none, black, dashed] coordinates {(0,2) (23851,2)};
	
	\addplot[mark=none, black, dashed] coordinates {(0,0) (0,3.1)};
	\addplot[mark=none] coordinates {(1606,3.1)} node [above] {\footnotesize Phase 1};
	
	\addplot[mark=none, black, dashed] coordinates {(3213,0) (3213,3.1)};
	\addplot[mark=none] coordinates {(10054,3.1)} node [above] {\footnotesize Phase 2};
	
	\addplot[mark=none, black, dashed] coordinates {(16895,0) (16895,3.1)};
	\addplot[mark=none] coordinates {(20373,3.1)} node [above] {\footnotesize Phase 3};
	
	\addplot[mark=none] coordinates {(11925,2.5)}  node [midway,draw,rectangle, minimum width=2cm] {\footnotesize Participant 1};
	

	
	\addplot [
	mesh,
	line width=1.2pt,
	scatter, mark=none,
	scatter src=explicit symbolic,
	shader=flat corner,
	point meta=\thisrow{color}
	] table [col sep=semicolon] {figures/graphs/Kirby.csv};

	\end{groupplot}
	\end{tikzpicture}
	\caption{
		Progress visualization of the expert (top), P8 (middle), and P1 (bottom). The $x$-axis shows time in minutes, and the $y$-axis measures progress based on the progress metric described in Section~\ref{subsubsec:behavanalysis}. Blue lines indicate manual interactions with the netlist, while green lines indicate interactions with the netlist via semi-automated scripts. Reaching the horizontal lines~1, 2, 3 marks completion of the respective phases, which are also separated by vertical dotted lines.}
	\label{fig:strategy2}
\end{figure*}

\end{document}